# Fabrication of 3D Nanoporous Gold with Very Low Parting Limit Derived from Au-based Metallic Glass and Enhanced Methanol Electro-oxidation Catalytic Performance

*Yi Xu, Dr. Pak Man Yiu, Dr. Guangcun Shan,* Prof. Dr. Tamaki Shibayama, Prof. Dr. Seiichi Watanabe, Prof. Dr. Masato Ohnuma, Prof. Dr. Wei Huang, Prof. Dr. Chan-Hung Shek*$^{*}$*

Yi Xu, Dr. Pak Man Yiu, Dr. Guangcun Shan, Prof. Chan-Hung Shek
Department of Physics and materials science, City University of Hong Kong, No.83 Tat Chee Avenue, Kowloon Tang, Hong Kong, China
Email: apchshek@cityu.edu.hk.

Dr. Guangcun Shan.
School of Instrumentation Science and Opto-electronics Engineering, Beihang University, No.37 XueYuan Road, Beijing 100095, China
Email: gshan2-c@my.cityu.edu.hk.

Prof. Tamaki Shibayama, Prof. Seiichi Watanabe
Laboratory of Quantum Energy Conversion Materials, Centre for Advanced Research of Energy and Materials, Division of Quantum Science and Engineering, Faculty of Engineering, Hokkaido University, Kita 13, Nishi 8, Kita Ku, Sapporo 0608628, Japan

Prof. Masato Ohnuma
Laboratory of Quantum Beam System Engineering, Division of Quantum Science and Engineering, Faculty of Engineering, Hokkaido University, Kita 13, Nishi 8, Kita Ku, Sapporo 0608628, Japan

Prof. Wei Huang
Institute of Advanced Materials (IAM), Jiangsu National Synergetic Innovation Centre for Advanced Materials (SICAM), Nanjing Tech University 30 South Puzhu Road, Nanjing 211816, China

**Keywords**: nanoporous gold, metallic glass, metal migration, methanol electro-oxidation;

**Abstract**

Nanoporous gold (NPG) with bi-continuous ligaments and pores structure has promising potential in functional applications, among which one prominent example is fuel cell electrocatalyst. However, current application of NPG is mostly limited to methanol electro-oxidation due to its weak catalytic performance. Here we report a simple chemical dealloying process, for generating peculiar three-dimensional (3D) free-standing NPG with high specific surface area associated with a novel porous 'cone shaped protrusion' morphology. This NPG structure possesses the highest specific activity of catalytic performance reported NPG catalysts so far.

1. **Introduction**

Nanoporous metals, such as gold, palladium, platinum, have been attracting increasing attention for a wide variety of applications in functional areas[1-2], since the porous structure exhibit excellent performance in applications like catalysis[5], surface enhanced Raman scattering (SERS)[6-8], supercapacitors[9] and actuators[10]. Porous structure is especially ideal for catalytic application because of the large surface-to-volume ration as well as the interconnected channel and metal ligaments, which accelerate the electron and molecules transportation, and hence excellent electrical conductivity. Among these nanoporous metals, nanoporous gold (NPG) has achieved more attention and several approaches, such as chemical dealloying and electrochemical dealloying techniques were developed to fabricate NPG.[3-4] There are two key principals for designing precursor materials and dealloying processes to fabricate NPG, namely, the "parting limit" or "dealloying compositional threshold" and the critical potential for electrochemical dealloying.[11] "Parting limit" is correlated with percolation threshold $P_c$. If the atomic fraction of the more reactive component is below $P_c$, the dealloying process does not occur.[12-14] The prototypical binary alloy system as precursor for NPG is Ag-Au. The experimentally determined parting limits for Ag-Au alloys are often higher than the percolation threshold, and are approximately 55 at % of Ag.[12,15,16] Besides the binary alloy system, fabrication of NPG with multicomponent alloy was also studied, such as Ag-Au-Pt ternary alloy and Au-Ag-Pd-Cu-Si amorphous alloy.[17,18] However, the fraction of less-noble elements being dissolved is still higher than 55%, NPG fabricated from precursor with active component content less than the percolation threshold 55% has never been reported up till now.

NPG, being an electrocatalyst for methanol oxidation in alkaline media, is expected to play an important role in full cell applications. It is because the activity of gold in alkaline media is much higher than that of platinum and the porous structure possesses more active sites arising from lattice strain effect and atomic sites on surface steps and kinks with low coordination number.[19,20] However, the structural features of NPG tend to coarsen during methanol oxidation cycles.[21] This shortcoming motivated some alternative approaches, for instance, the fabrication of Au-based bimetallic nanoporous structures containing real catalyst Platinum or Palladium by dealloying ternary alloy precursors[22], or plating Pt onto NPG by electrodeposition[23], or synthesizing nanoporous structure with hollow ligaments by galvanic replacement reaction (GRR) of other metal with Pt ions precursor.[24] However, since Au is miscible with Pt or Pd in bulk, therefore in order to improve the binding reaction between metal surface and molecules, the ligand effects are used to alter the electronic structure of the interface between these two metals in addition to the ensemble effect at nanoscale.[25-26] While the electronic interaction induced by metal migration on Pd-Au bimetallic alloy particle with core-shell structure has been reported [27], the influence of electronic interaction between nanoporous Au and Pd thin film has not been reported in the literature so far.

Au-Cu-Si ternary metallic glass (MG) was chosen as the precursor to fabricate NPG in this investigation and NPG could be fabricated without the constraint of the limiting principal. The atomic fraction of the less noble metal, copper in this case, was below 25%, therefore resulting in appreciable saving in the cost compared with that of silver content exceeding 55% for Au-Ag binary alloy system. In the initial study, we show the peculiar topology with porous 'cone shaped protrusion' structure generated on the top of NPG and analyze the formation mechanism through HADDF-STEM techniques. This specific cone-shaped morphology has never been observed from dealloying normal binary alloy precursor. Owing to the three-dimensional (3D) porous 'cone shaped protrusion' morphology, a relative large specific surface area and a high concentration of low-coordination atomic sites were generated. This structure possesses superior methanol oxidation reaction (MOR) catalytic performacne compared to prototypical nanoporous structure in previous researches[28,29]. In addition, we chose NPG after dealloying 30 mins as a platform to fabricate NPG@Pd nanocomposite catalyst. Electro-chemically chronoamperometry technique was used to dissolve Co from $Pd_{50}Co_{50}$ thin film deposited on NPG. The resulting NPG@Pd catalyst presents very good catalytic methanol oxidation performance and long-term stability against electrochemical cycling. The enhancement of MOR attributed to the interplay of the porous 'cone shaped protrusion' structure with more catalytic active sites and the electronic interaction through a combination of migration of Au into Pd layer and Co dissolution.

2. Results and Discussion

2.1. Fabrication Process

The fabrication strategy shown in **Figure 1** is to combine the chemical dealloying with thin-film-deposition/electrochemically-dealloying processing routes, through which the preparation of NPG and attachment of Pd to ligaments channel are conducted successively.

Magnetron sputtering was used to coat $Pd_{50}Co_{50}$ thin film on the surface of the NPG fabricated from as-spun $Au_{55}Cu_{25}Si_{20}$ metallic glass ribbon with $T_g$ of 90ºC (Figure S1, Supporting Information) by chemical de-alloying in iron chloride solution at 80ºC. Co was electrochemically dissolved away and that induced metal migration between the Pd film and NPG, since Au atoms with low surface energy was prone to diffuse to surface and change the electronic structure by displacing Pd atoms. That outward diffusion of Au also leads to a rougher surface and thus a larger number of active sites were generated by leaching out of Co.

2.2. Characterization and formation-mechanism analysis of NPG

2.2.1. Structural characterization of NPG

The SEM micrographs in **Figure 2** show the planar view (Figure 2 a-c) and cross section morphology (Figure 2d) of the NPG fabricated by dealloying $Au_{55}Cu_{25}Si_{20}$ MG ribbon. Nanoporous structure of more than 10μm thick formed on both sides of the ribbon. The nanoporous region consisted of channels and ligaments with average size of ~50 nm. The composition of the nanoporous region, measured by EDS (Table S1, Supporting Information), confirmed that only copper was leached out during the dealloying process. Interestingly, "cone shaped protrusions" were observed (Figure S2a-c, Supporting Information) on the surface of the nanoporous ribbon and the density of the "cone shaped protrusions" increased with increasing dealloying time.

The cross-section views of the dealloyed ribbon (Figure 2d and Figure S2d, Supporting Information) show the uniform porous structure within the "cone shaped prostrusions" areas, which contributed to the high specific surface area. By using Brunauer-Emmett-Teller (BET) technique (Figure S3a-b, Supporting Information), the specific surface area of samples after 30 mins and 60 mins dealloying are 18 $m^2/g$ and 31 $m^2/g$, respectively. X-Ray Diffraction (XRD) patterns (Figure S4a, Supporting Information) of samples obtained after dealloying for 3, 10 and 30 mins, denoted as D-3, D-10 and D-30 respectively, indicate that the amorphous ribbon crystallized after dealloying. The intensity of the amorphous hump corresponding to the glassy structure decreased progressively. The D-30 specimen shows sharp diffraction peaks at 2θ values of 38.2°, 44.4°, 64.6°, 77.6° and 81.7°, which correspond to the face-center-cubic (FCC) phase of metallic gold (JCPDS, card No. 04-0784).

### 2.2.2. Formation mechanism study of NPG.

In order to study the formation mechanism of the "cone shaped protrusion" nanoporous structure on the MG ribbon surfaces after dealloying, TEM samples of D-30 were prepared with Focused Ion Beam (FIB). The samples were examined with a spherical-aberration-corrected TEM in STEM mode with a high-angle annular dark-field (HAADF) detector (**Figure 3)**. Figure 3a shows a bright-field HAADF-STEM image of the overview of one of the 'cones'. Figure 3b-c show high-resolution images located at the top region (Figure 3b) and at the curve bottom region (Figure 3c) of a porous 'cone shaped protrusion' area. Selected area electron diffraction patterns (SAED) were also collected and a HAADF-STEM image with composition x-ray mapping of Au-L$\alpha$, Cu-L$\alpha$, Si-L$\alpha$ lines respectively are shown.

In the D-30 sample, nanoporous structure was generated on the metallic glass ribbon surface and some porous "cones" protruded from the ribbon surface after surface dealloying stage. Three-dimensional bi-continuous porous structure with interspersed positive curvature ligaments and negative curvature pores was formed. The sizes of pores and ligaments near the tip of the 'cone shaped protrusions" (region 'b' in Figure 3a) were larger relative to those near the bottom (region 'c' in Figure 3a). It is proposed that the continuous dissolution into bulk generated at specific site, in which MG state turned into crystalline grain, and the initial nanoscale pores and ligaments formation in region 'c' grew up normal to curve shaped grain boundary. This proposed mechanism is supported by the diffraction pattern (inset Figure 3c and Figure S5a, Supporting Information) with the [010] zone axis, which was obtained throughout the curve bottom region. This indicated that the whole nanoporous structure generated at region 'c' was a single crystal or consisted of grains with negligible misorientation. The diffraction patterns were indexed for gold silicide with orthorhombic unit cell. The lattice constants, *a = 0.95nm, b = $\sqrt{2a_{Si}}$ = 0.76nm, c = 0.67nm* (Figure S5b-c, Supporting Information), agreed with the reported result. [30]

At the final dealloying stage, the ligaments thickness grew with the pores size. In Figure S6a-b (Supporting Information), high resolution TEM micrographs showed that the ligaments consisted of nanoparticles surrounded by amorphous structure. The lattice plane (Figure S6c, Supporting Information) separation of nanoparticle was measured as 0.235nm, corresponding to Au {111} planes and the amorphous structure was $SiO_x$. This agrees with the HAADF-STEM composition mapping which suggested that silicon tended to diffuse to the top surface of the ligaments. Figure S6d (Supporting Information) proves the presence of amorphous Silicon-Ox phase and gold nanoparticle with step feature and Au (220) plane on the surface of a ligament.

It is proposed that the generation of small nanoscale ligaments in the initial porosity formation stage at region 'c' may lead to the volume expansion to generate the porous 'cone shaped protrusion'. Because the nanoscale ligaments with high positive and negative curvature leads to large tension and compression strain state respectively. [19] In the pore development stage, the ligaments size increases with the decomposition of gold silicide complex into Au clusters and $SiO_x$ mixture due to the metastability of gold silicide. In the final stage, gold clusters aggregated to Au particle and the remaining Si diffused out to surface and oxidized into $SiO_x$ eventually in the final coarsening stage at region 'b'.

### 2.3. Characterization and metal migration effect analysis of NPG@Pd
#### 2.3.1. Structural characterization of NPG@Pd catslyst

Nanoporous palladium on NPG substrate (NPG@Pd) catalyst was synthesized by magnetron sputtering/electro-chemical dealloying methods, in which the thickness and morphology of Pd layer can be controlled by deposition time and the time of dealloying. The optimized sputtering/dealloying condition was found to be a combination of deposition to 100nm Pd-Co film thickness followed by 500 s electro-chemical dealloying, through which the Co was almost completely dissolved. In order to verify the migration of Au atoms into Pd layer with the increase of dealloying time, three sets of samples, namely, the ones dealloyed for 500 s, 1000 s and 2000 s, respectively, were examined in detail. The three samples were denoted as 'NPG@Pd 500', 'NPG@Pd 1000' and 'NPG@Pd 2000', respectively, where the number signified the electro-chemical dealloying time.

Figure 2e illustrates the plane-view of the NPG after deposition of Pd-Co thin film by magnetron sputtering. The formation of Pd-Co thin film sealed the flat ribbon surface (Figure 2e, inset) so that nanoporous gold structure formed in the prior dealloying could not be seen. [31] After dealloying in sulfuric acid at 0.6V (vs SCE), the NPG structure was revealed again in the three NPG@Pd samples (Figure 2f-h). The cross-section view of area between 'cones' (**Figure 4**) shows the interface between the NPG and the Pd-

Co thin film after electro-chemical dealloying. In Figure 4a, STEM image shows the porous thin film layer on the top of aggregation hillocks after electrochemical dealloying. The dealloyed thin film formed a continuous nanoporous structure, chemically bonded with the surface of gold-rich substrate, and offered excellent electrical conductivity between Au and Pd.

### 2.3.2. Metal migration effect analysis of NPG@Pd catalyst

The schematic diagram in Figure 1 illustrates that electro-chemical dealloying resulted in selective dissolution of Co and migration of Au to the Pd-Co thin film simultaneously. This was confirmed by X-ray photoelectron spectroscopy (XPS) depth profile in sample 'NPG@Pd 500'. Figure 4c illustrate that XPS peak of Au 4f region shifted to higher binding energy (BE) and the intensity increased with the increasing depth, which agreed with the result from STEM-EDS line scan (Figure 4b), indicating Au atoms tended to migrate into the Pd-Co thin film and the sample surface. Meanwhile, Figure S7 (Supporting Information) shows the XPS peak intensity of Co 2p decreased to zero when approaching the interface between the thin film and the NPG. Co atoms were almost completely dissolved from the Pd-Co thin film after dealloying for 500 s, except that about 5 at.% Co still remained on the top surface of the thin film in the form of metallic $Co^0$ and valance state $Co^{2+}$ as shown in the inset of Figure S7 (Supporting Information).

In order to confirm the only existence of Au atoms migration into Pd thin film layer, the composition and valence state of the NPG and 'NPG@Pd' were further examined by core-level X-ray photoelectron spectroscopy (XPS), shown in **Figure 5**.

In the Figure 5a, the spectra of Au 4f region from NPG was fitted with two Gaussian peaks. It exhibited an intense Au $4f_{7/2}$ (Au $4f_{5/2}$) peak at 84.7 eV (88.4 eV), which is 0.8eV higher in binding energy (BE) than that of bulk metallic Au but only 0.3 eV lower than that of gold silicide. [32,33] It suggested that the NPG structure was a mixture of metallic gold and gold silicide. It could be confirmed from the spectra of Si 2p region (Figure S8, Supporting Information), which is in accordance with the previous result.[34] The main peak appearing at 103.0 eV with an energy shift of 3.7 eV, compared to that of bulk Si at 99.3eV, could be assigned to $SiO_2$. In addition, there was an ambiguous weak peak at 100.4 eV, which was approximately corresponding to sub-oxide Si oxidation state. After electrochemical dealloying induced metal migration, Figure 5b-d shows the spectra of Pd 3d region from the three 'NPG@Pd' samples after fitting with four Gaussian peaks in the spectrum. Their peak profiles became sharper and more symmetric compared to those before electrochemical dealloying shown in Figure 5f. It is noted that double intense peaks for Pd $3d_{5/2}$ and Pd $3d_{3/2}$ were evidently featured at 335.6 eV and 340.9 eV, separated by ~5.3eV, indicating the metallic Pd state. Two weak shoulder peaks for Pd $3d_{5/2}$ and Pd $3d_{3/2}$ were also observed at 337.1 eV and 342.4 eV and they could be related to PdO or Pd silicide, both of which possessed similar BE positions. However, indication of Pd silicide formation could not be seen from the Si 2p spectra, which remained nearly constant at 99.3eV from metallic Si. Since the average position of Si 2p peak from the three 'NPG@Pd' samples were still expected to be around at 103eV. Moreover, the absence of distinctive O 1s peak in the Pd $3p_{3/2}$ and O 1s region (shown in Figure 5g) was attributed to the overlapping with strong Pd$3p_{3/2}$ peak from metallic Pd located at 532.6 eV. It was reported that Pd $3p_{3/2}$ peaks were located at 533.7 eV for Pd silicide and O 1s peak at 530.2 eV for PdO.[35] In the present work, there is no Pd silicide peak at 533.7eV to verify the presence of Pd silicide at a quantifiable component. For Pd $3p_{1/2}$ region, peak at 560.5eV still corresponded to metallic Pd.

In addition, it was shown that all three NPG@Pd samples showed the shift of Au 4f7/2 peaks to lower BE at 83.8 eV, which was 0.1 eV lower than that of metallic Au and considerable increase in peak intensities, which could be an indication of migration of Au atoms. XRD results (Figure S4b-c, Supporting Information) also showed shifts of the Au (111) plane peak from 38.1° for nanoporous metal to 38.2 for 'NPM@Pd 500' and 'NPM@Pd 1000' samples, while the Au(111) peak for 'NPM@Pd 2000' sample was located at 38.1°, which may be attributed to the high atomic percentage of Au after longer diffusion time.

### 2.4. Electro-chemical performance
### 2.4.1. Electro-chemical performance of NPG

In order to understand the correlation between the catalytic performance and this specific porous structure, we examine the methanol oxidation reaction (MOR) performance (**Figure 6**a) of as-cast Au-based MG ribbon and NPG after dealloying for 15 mins and 30mins, which are denoted as D-15 and D-30 respectively, in 0.5M KOH

+ 1M CH$_3$OH solution, shown in Figure 6a.

In the positive potential sweeping scan of D-30 sample, the onset potential of methanol oxidation occurred at 0.03 V. The anodic current density reached a maximum at 0.20V (vs. SCE), which corresponded to the oxidation of methanol. At higher potential (> 0.5V), the sharp arise of current is associated with the reactivation of the anodic oxidation of methanol by gold oxides.

In the negative sweeping scan, the reduction peak observed at 0.04 V is attributed to the reduction of Au-Ox. A second oxidation peak at -0.07V in the reverse scan is ascribed to the removal of the incompletely oxidized carbonaceous species formed in the forward scan and Au-Ox reinitiate the methanol oxidation.

Interestingly, the maximum specific activity of as-spun Au-based MG ribbon is ~ 93μAcm$^{-2}$, which is higher than traditional NPG electrode obtained by dealloying of Au-Ag alloy (~88 μAcm$^{-2}$).[28] With dealloying for 15 mins to 30 mins, the current density was increased to ~ 120 μAcm$^{-2}$ and ~136 μAcm$^{-2}$, respectively, which were both higher than that of NPG nanoparticles (~117μAcm$^{-2}$).[29] In addition, the long-term stability of these catalytic electrode was evaluated by chronoamperometry curves as shown in Figure 6b. The polarization current density decayed initially and reached a stable value after being polarized at 0.29 V (vs SCE) for ~1000 s. D-30 exhibited a better catalytic stability after ~1500 s compared with D-15, indicating the enhanced catalytic stability of the porous 'cone shaped protrusion'.

It was observed in Figure S6c-d (Supporting Information) that the top region of the porous 'cone shaped protrusion' composed of curved ligament and Au nanoparticles on the surface of ligaments, which contain high density of low-coordination atoms as ideal catalytic active sites. The porous structure provides a high density of intrinsic catalytic sites, due to the large curvature gradients between concave and convex regions which geometrically require steps and kinks with a high concentration of low-coordination atomic sites.[20]

On the other hand, theoretical and experimental works[36,37] have shown that elastic strain can enhance molecular interaction and hence reduce energy barriers for chemical reactions. Thus, the interplay of porous structure and strain effect inside porous 'cone shaped protrusion' was effective for enhancement of catalytic ability.

### 2.4.2. Electrochemical performance of NPG@Pd catslyst

Au atoms migration into the Pd layer was confirmed from XPS results discussed above, so it is essential to study the relationship between methanol electro-oxidation reaction (MOR) and the effect of Au atoms migration.

The methanol electro-oxidation reaction (MOR) of NPG@Pd samples was tested in 0.5M KOH + 1M $CH_3OH$ solution. Alkaline solution was chosen because methanol electro-oxidation is more active in alkaline than in acidic solution. More importantly,

OH$^-$ ions do not facilitate Pd dissolution, which may happen in the presence of sulphate ions if sulphuric acid is used.[38]

Figure 6c indicated the existence of Au migration enhanced MOR catalytic reaction, which is confirmed by higher MOR activity of 'NPG@Pd 2000' compared with that of sample 'NPG@PdCo' without de-alloying and the bare NPG sample. Moreover, the increase of electro-chemical dealloying time induced metal migration enhanced MOR catalytic performance shown in Figure 6d. Methanol oxidation occurred in different potential regions. In the positive scan, the Region I at lower potential ranging from -0.8V to 0.2V was the characteristic of methanol oxidation on Pd and overlaps with Au oxides formation. Moreover, there was also an observable weak peak (indicated by arrow in Figure 6d) between a broad methanol oxidation peak on Pd (Region I) and sharp increase of the peak of methanol oxidation on Au at higher potentials (>0.3V, Region II), which was also associated with the reactivation of methanol oxidation by Au oxides. In the alkaline solutions containing methanol, OH$^-$ ions are easily trapped on the surface of nanoporous metal to facilitate the generation of Au oxides and methanol could then be oxidized on these surface Au oxides at higher positive potential scan (region II). Reduction of Au oxides took place at ~0V (region III) followed by the methanol oxidation peak (region IV) in negative potential scan induced by the removal of the incompletely oxidized carbonaceous species and Au oxides formed in the positive potential scan.

In addition, the onset oxidation potential, peak potential and specific activity, normalized by ECSA, are summarized in supplementary Table S2 (Supporting Information), showing the onset potential for methanol oxidation shift negatively from -0.324V to -0.402V, which is more negative than that of Pt/C commercial catalyst. The specific activity for the three NPG@Pd samples were 1.76 mA/cm$^2$, 1.32 mA/cm$^2$ and 2.14 mA/cm$^2$ respectively, which is also higher than that of Pt/C commercial catalyst, indicating higher catalytic ability for methanol oxidation reaction due to Au migration to the Pd layer.

To meet the requirements for fuel cell applications, we also evaluated the stability through multiple electrocatalytic cycles of methanol oxidation at 100 mV/s, shown in **Figure 7**a-c. After 1000 cycles, the oxidation peak current density, which represents electrocatalytic performance, for the three NPG@Pd samples retained 75, 90 and 85%, respectively, of their initial values. All three samples exhibited high stability and no significant deactivation, especially for sample 'NPM@Pd 2000'. No obvious change in the cyclic voltammetry(CV) profile (Figure S10b, Supporting Information) could be observed after 1000 cycles, indicating enhance catalytic stability due to bimetallic interaction between Au and Pd.

To explain this trend in MOR activity, we focused on sample 'NPG@Pd 2000' which has more extensive Au migration and exhibited good catalytic ability. Alcohol oxidation reaction is more active in alkaline media than in acidic media, and alcohol

tends to deprotonate at higher pH value:

$$H_\beta R - OH_\alpha \rightarrow H_\beta R - O^- + H_\alpha^+ \quad (1)$$

$H_\alpha$ and $H_\beta$ represent alpa-Hydrogen elimination and beta-Hydrogen elimination respectively. The oxidation activity of R and pKa (Acid dissociation constant) are dependent on Hammett-Type relationship, which states that low pKa value leads to higher reactivity. Nevertheless, methanol, having a low oxidation activity, is an exception to this rule, which is attributed to the stronger methanolic C-$H_\beta$ bond than other alcohols. In the equation above, the first deprotonation, viz. the alpa-hydrogen elimination, is not influenced by the 'real' catalyst but by the pH value of the solution, due to O-H bond activation. Therefore, alpa-hydrogen elimination by the metal alone is unlikely.[39]

The remaining alkoxide after initial alcohol oxidation step is thought to be more active toward oxidation and therefore readily oxidized to aldehyde:

$$H_\beta R - O^- \rightarrow R = O + H_\beta^+ + 2e^- \quad (2)$$

For methanol with a relatively strong C-$H_\beta$ bond, a catalyst is necessary for the $H_\beta$ dehydrogenation and the presence of catalyst becomes crucial. For pure Au catalyst, Bhushan N. Zope[40] suggested that the presence of adsorbed OH⁻ on gold surface lowers the barrier for the subsequent β-hydrogen elimination to form aldehyde, indicating the interaction between the alkoxide and the gold surface is mandatory.

In a nutshell, Au has no essential role for α-hydrogen elimination, but provided assistance for β-hydrogen elimination, acting primarily as an electron acceptor. Additionally, the aldehyde is not stable in alkaline media and decompose quickly, which determine the final products of oxidation reaction. The activity of gold in alkaline media can be very high, even higher than that of platinum. It is because Au as catalyst does not form poisoning intermediates on the surface and Au surface can oxidize CO-like intermediate species to refresh the active sites by producing $CO_2$. [41]

Therefore, we can attribute the enhancement of catalytic ability to the following reasons:

a) The dissolution of Co atoms from $Pd_{50}Co_{50}$ thin film deposited on NPG results in the rough surface and increase of active sites. This proposed mechanism agrees with the observation that the relatively poor MOR catalytic ability for $Pd_{50}Co_{50}$ thin film on NPG was substantially improved after the leaching of Co atom and exposing the nanoporous structure underneath.

b) The metal migration effect between Pd and Au during dissolution process modified the electronic structure of Pd with Au addition which lowers the barrier for the β-hydrogen elimination and formation of CO-like oxide intermediate species to release the active sites by producing $CO_2$. Meanwhile, the interplay of low-coordinated chemical active sites and strain effect inside porous 'cone shaped protrusion' areas also promoted the catalytic performance.

3. **Conclusions**

We have developed a simple method to fabricate 3D bicontinuous NPG by selective leaching of electrochemically more active elements from the starting material. The formation of 'cone shaped protrusion' nanoporous structure on the surface of the NPG gave a high specific surface area of $31 m^2 g^{-1}$ after chemical dealloying of Au-based MG with a low 'parting limit' of 25% (lower than theoretical 'parting limit' 55%). Morover, the fabricated NPG delivered superior specific activity value compared with reported NPG catalysts. This excellent performance was attribute to the porous 'cone shaped protrusion' consisting of curved ligament and Au nanoparticles on the surface of ligaments, which contain high density of low-coordination atoms as ideal catalytic active sites. In addition, based on the NPG feature, a NPG@Pd catalyst was fabricated with magnetron sputtering followed by electrochemical dealloying and greatly enhanced MOR performance was achieved. The enhancement of MOR activity can be attributed to **a**) the rough surface and increase of active sites due to Co dissolution by electrochemical dealloying; and **b**) Au atoms metal migration that modify the electronic structure of Pd to lower the barrier for the β-hydrogen elimination and CO-like oxide intermediate species to release the active sites. With these combined benefits, the method of fabricating NPG and NPG@Pd catalyst are expected to have an impact on the design of future electrocatalysts for fuel cell development.

4. **Experimental Sections**

*Nanoporous gold (NPG) sample preparation*: $Au_{55}Cu_{25}Si_{20}$ MG ribbon with approximate 50μm thickness was prepared by melt-spinning technique. A single rotating Cu roller was used for melt-spinning. Free-standing NPG ribbon were fabricated by chemically dealloying the $Au_{55}Cu_{25}Si_{20}$ MG ribbon in iron chloride solution (Sigma-Aldrich) at 80ºC. The copper was dissolved selectively from the metallic glass ribbon. The as-prepared NPG ribbons were carefully rinsed with deionized water (18.2 MΩ.cm) to remove the residual iron chloride solution.

*Characterization*: Scanning electron microscopy (SEM) images were obtained with a JSM 6335F FE-SEM (JEOL) operating at 5 kV, Energy-dispersive X-ray spectroscopy (EDS) was recorded at 20 kV. For examining the cross-section of 'cone shaped protrusion' regions, TEM samples were obtained with a JEM-9320 Focused Ion Beam (FIB) system. HRTEM images were obtained with a JEM-2100F (JEOL) transmission electron microscope operated at 200 kV. HAADF-STEM images and elemental mapping was obtained with spherical aberration corrected Scanning Transmission Electron Microscope (FEI. Titan G2) Titan G2 60-300. XRD patterns were recorded on a Rigaku XRD diffractometer with Cu Kα radiation. The thermodynamic parameters of the glassy alloy were examined by using a Perkin-Elmer DSC7 at a heating rate of 0.67K/s. XPS spectra were measured using a PHI5802 system (Physical Electronics) with a monochromatic Al Kα X-ray radiation source. All XPS spectra were fit using XPS Peak 4. XRD patterns were recorded on a Rigaku XRD diffractometer with Cu

Kα radiation.

*Estimation of electrochemically active surface area(ECSA)*: The electrochemical behavior of NPG and NPG@Pd catalysts were characterized using cyclic voltammetry(CV) technique in 0.5M $H_2SO_4$ aqueous solution at room temperature. Pd and Au electrochemically active surface areas (ECSA) were calculated in order to normalize methanol oxidation current density, and the value was obtained by integration of the consumed charge for the reduction region of gold oxide on electrode and dividing by 212 $\mu C/cm^2$ and 450 $\mu C/cm^2$, respectively. Details are shown in supplementary Figure S9-11, Supporting Information.

*Electrochemical behavior measurement*: Electrochemical measurements were carried out using a CH Instruments 760E electrochemical workstation. Three electrodes setup in 0.5M KOH and 1M $CH_3OH$ solution as electrolyte was employed, with the working electrode being the as cast Au-based metallic glass ribbon, NPG after dealloying and fabricated NPG@Pd catalyst as free-standing electrode. The working electrode area immersed into electrolyte was 5mm×5mm. saturated calomel electrode (SCE) was used as reference and the counter electrode was a Pt sheet. The methanol oxidation catalytic performance of as-cast Au-based metallic glass ribbon, D-15 and D-30 products were measured in the form of cyclic voltammograms. The long-term test was measured via chronoamperograms applied at 0.2V (vs SCE). The methanol oxidation catalytic performance of NPG and 'NPG@Pd 500-NPG@Pd 2000' were measured in the form of cyclic voltammograms at a scan rate 50 mV s$^{-1}$ in deoxygenated solutions containing 0.5 M KOH and 1M CH3OH. Repetitive potential cycling stability test was recorded at a scan rate 100 mV s-1 after 3, 200 and 1000 cycles.

**Supporting Information**
Supporting Information is available from the Wiley Online Library or from the author.

**Acknowledgement**

The work in this paper was financially supported by a General Research Fund from Research Grants Council, Hong Kong SAR Government (Project number: CityU 11211114). This work is also partially funded by National key research and development program (no. 2016YFC1402504) by ministry of science and technology. All HAADF-STEM, and EDS mapping were supported by Centre for Advanced Research of Energy and Materials in Division of Quantum Science and Engineering, Faculty of Engineering, Hokkaido University. We gratefully acknowledge the support from Dr. Ruixuan Yu and Dr. Yanhua Lei in Hokkaido University. The Brunauer-Emmett-Teller (BET) data were measured by Mr. Haidong Bian in Dr. Yangyang Li's group. The electrochemical performance measurement was supported by Mr. Zijie Tang in Dr. Chunyi Zhi's group and we acknowledge the discussions with Dr. Chunyi Zhi and Dr. Zhiguang Zhu. We acknowledge Mr. Tak Fu Hung for the use of JEOL 2100F for collection of HRTEM images. We acknowledge Shanghai Synchrotron Radiation Facility for access to the BL14W1 X-ray Absorption Fine Structure Spectroscopy (XAFS) Beamline.

monoxide on the oxidation of alcohols on a gold catalyst. *Nat. chem.* **2012**, 4(3), 177.

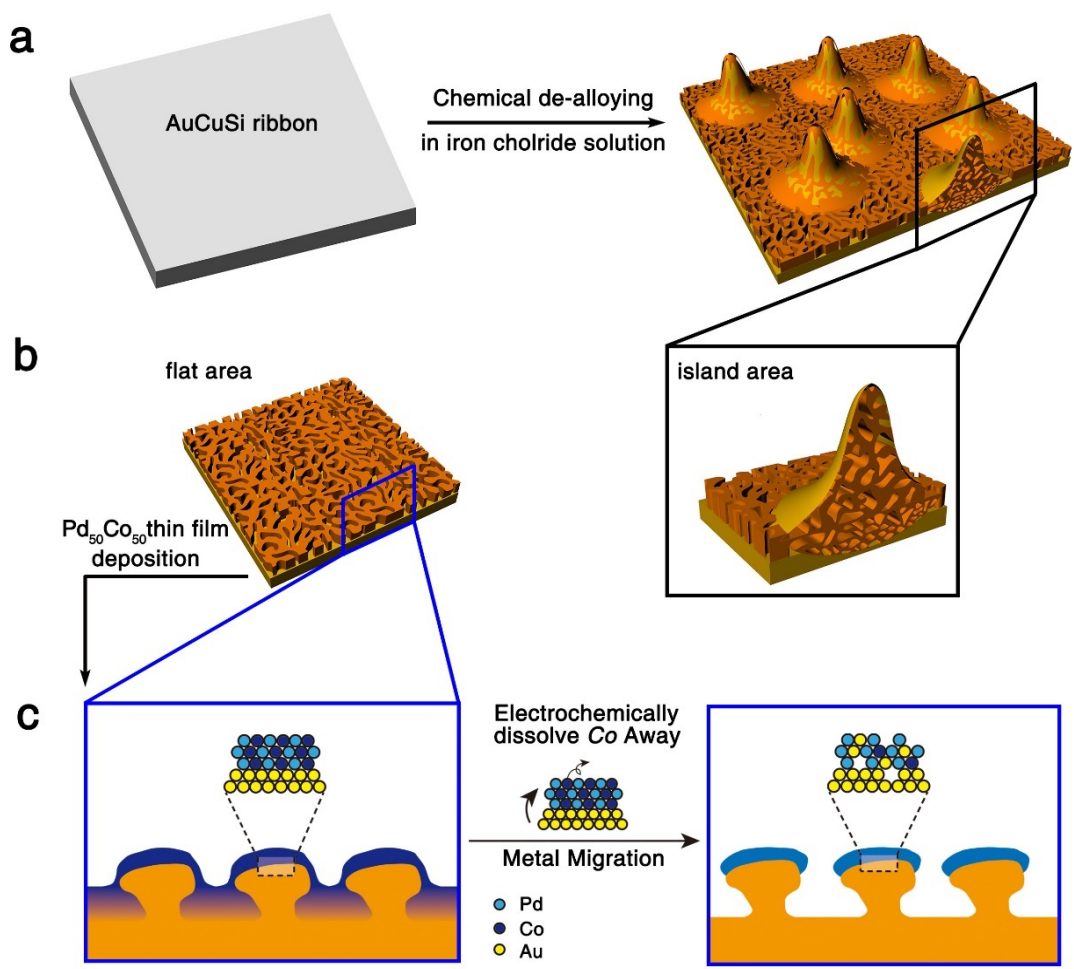

**Figure 1 schematic diagram showing the fabrication strategy for nanoporous gold (NPG) and NPG@Pd catalyst.** (a) Nanoporous gold was prepared by chemical de-alloying Au-based metallic glass ribbon. (b) 3D bicontinuous nanoporous gold with island areas and flat areas were formed. (c) Enlarged image selected from flat region shows the aggregation of hillocks, and NPG@Pd catalyst was synthesized by electrochemical dealloying deposited $Pd_{50}Co_{50}$ thin film, which induced Au metal migration and dissolution of Co element simultaneously.

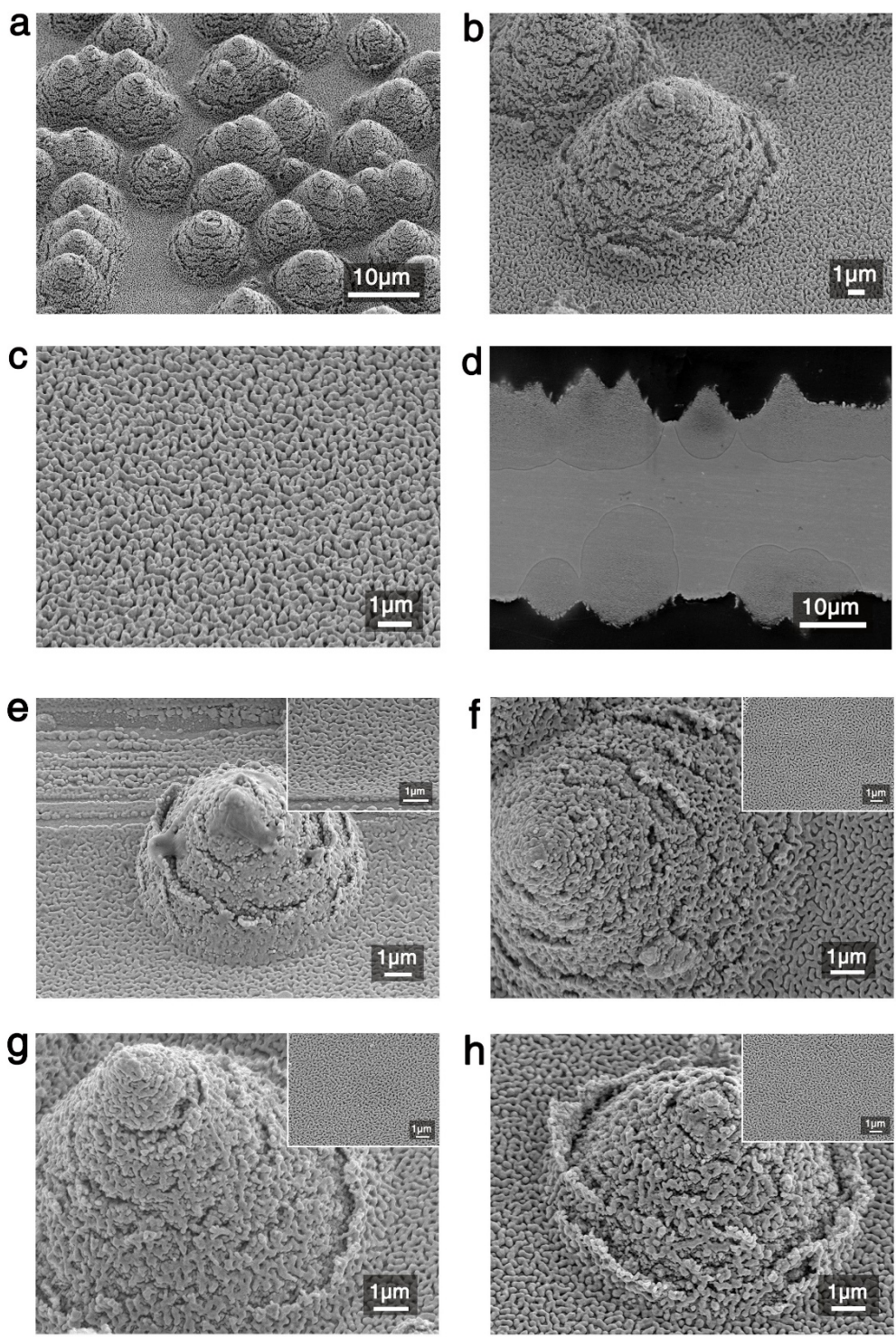

**Figure 2** (a) Plain view of high resolution SEM image of $Au_{55}Cu_{25}Si_{20}$ metallic glass ribbon after chemical dealloying 30mins. (b) Enlarged image for the single porous 'cone shaped protrusion' area and (c) the flat region area. (d) Cross-section view of low magnification SEM image for chemical dealloying 30mins. And high resolution SEM image of NPG@Pd catalyst before electrochemical dealloying (e) and after electrochemical dealloying (f) 500 s, (g) 1000 s, and (h) 2000 s in 'cone shaped protrusion' region and flat region(inset) respectively.

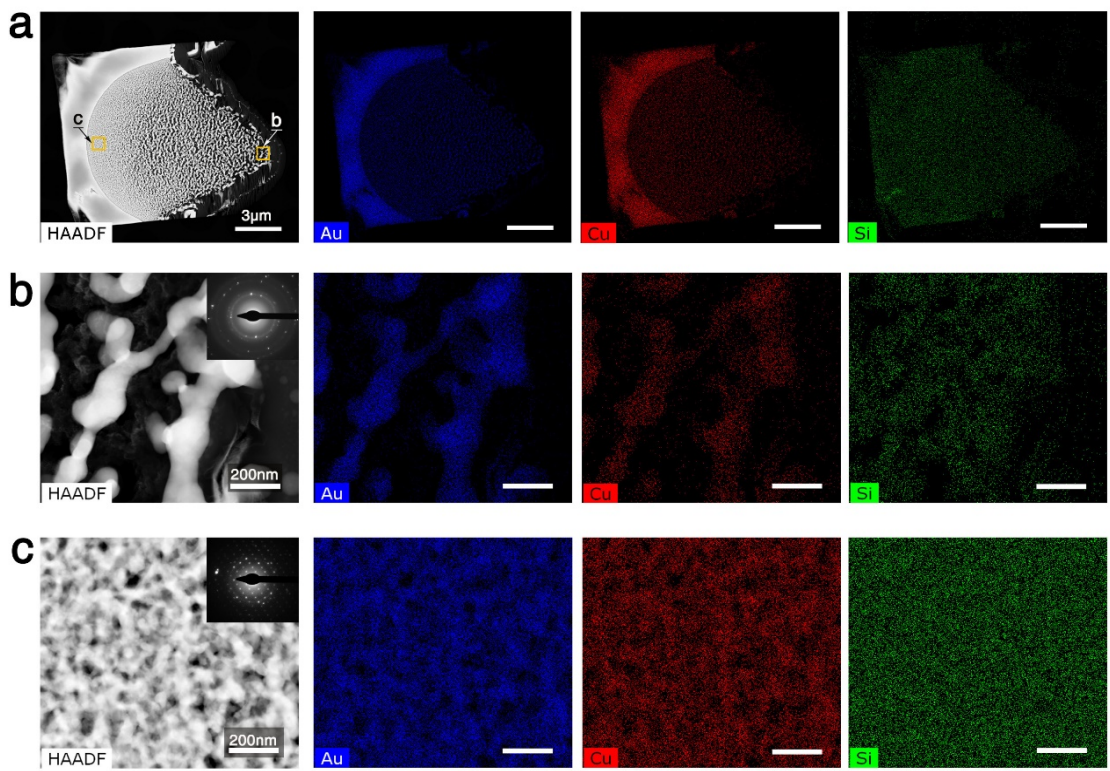

**Figure 3** Internal nanoporous structure characterization of focused ion beam (FIB) milled NPM by spherical-aberration-corrected TEM. (a) High-angle annular dark-field STEM images of 'cone shape' area and corresponding EDS elemental mapping; (b,c) Enlarged HADDF-STEM image and corresponding EDS elemental mapping from the selected area in a, selected area diffraction pattern (SAED) of corresponding HADDF-STEM image (insets in b and c).

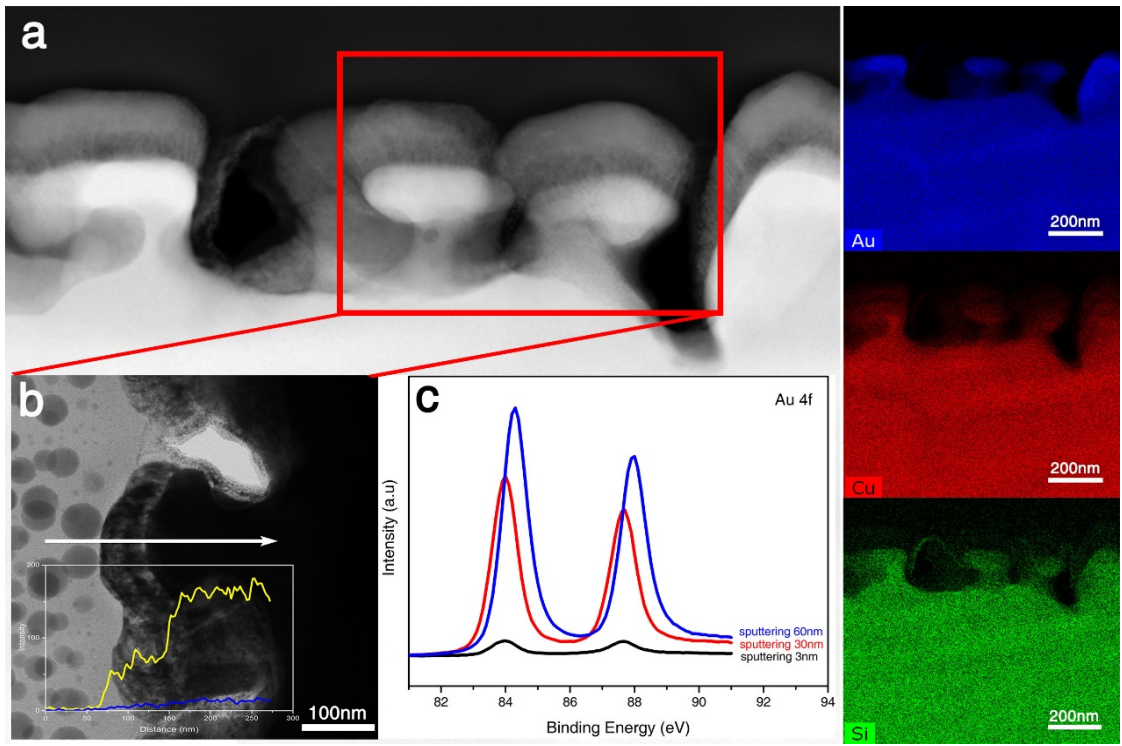

**Figure 4** (a) cross-section view of HADDF-STEM image and corresponding EDS elemental mapping in the right column from FIB milled 'NPM@Pd 500'; (b) STEM image and the intensity plots of line scan based on a raw EDS signal that obtained from the trace line is shown in the inset, showing Pd (blue color) and Au (yellow color) element distribution; (c) XPS depth profile of Au 4f region for sample 'NPM@Pd 500' after sputtering thickness with 3nm, 30nm and 60nm.

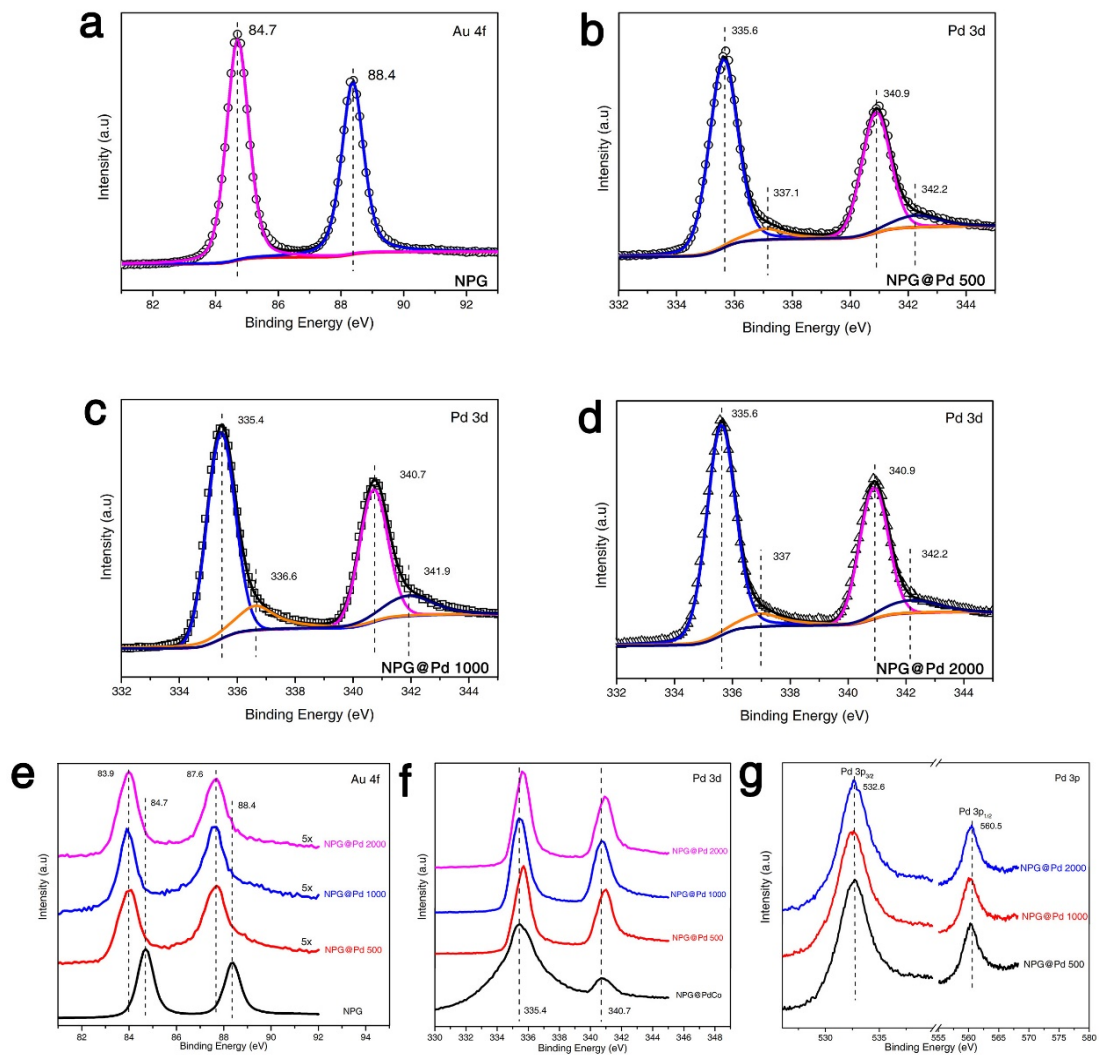

**Figure 5** (**a**) XPS spectra of the Au 4f region for NPG and Pd 3d region for (**b**) 'NPG@Pd 500' (**c**) 'NPG@Pd 1000' and (**d**) 'NPG@Pd 2000' after fitting, compared to (**e**) the Au 4f region for 'NPG@Pd 500-NPG@Pd 2000' and (**f**) the Pd 3d region before electrochemical dealloying respectively; (**e**) XPS spectra of the Pd 3p and O 1s region.

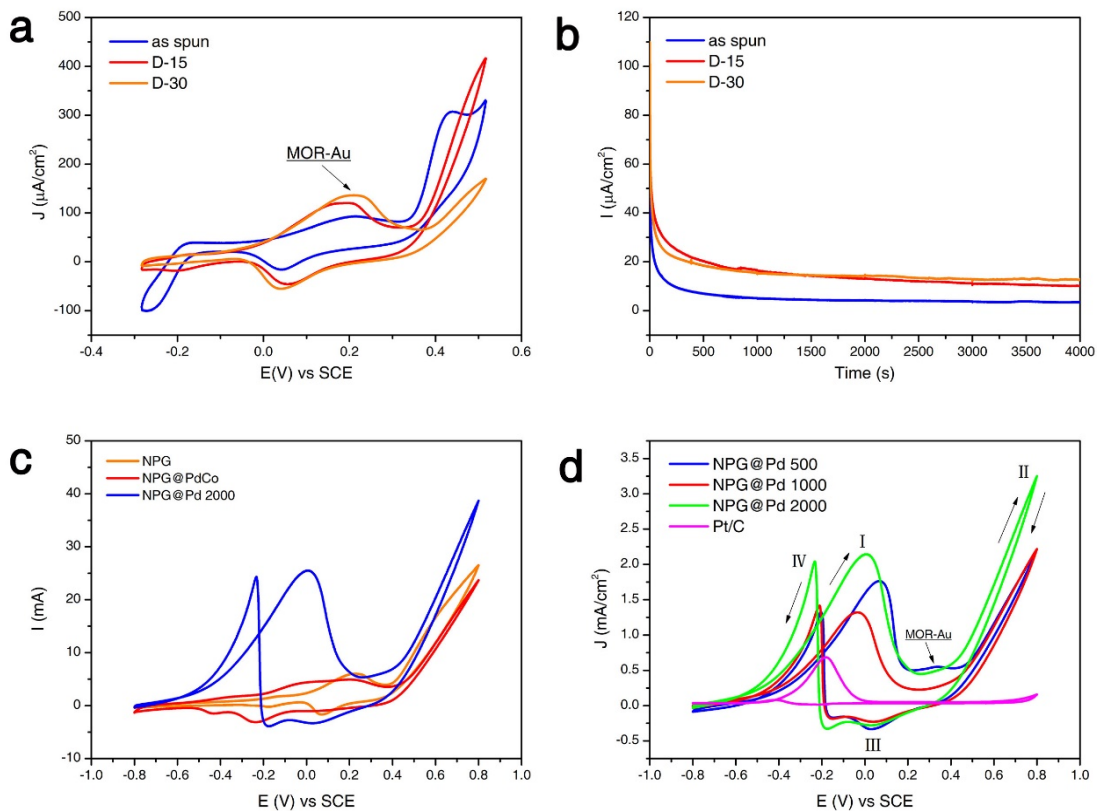

**Figure 6** The methanol oxidation reaction measurements via (**a**) cyclic voltammograms and (**b**) chronoamperograms of as-cast Au-based metallic glass ribbon, D-15 and D-30 products, and (**c**) that of 'NPG@Pd 2000' contrast to D-30 and NPG@PdCo without Electrochemically dealloying. (**d**) The methanol oxidation catalytic performance of 'NPG@Pd 500-NPG@Pd 2000' compared to commercial Pt/C catalyst.

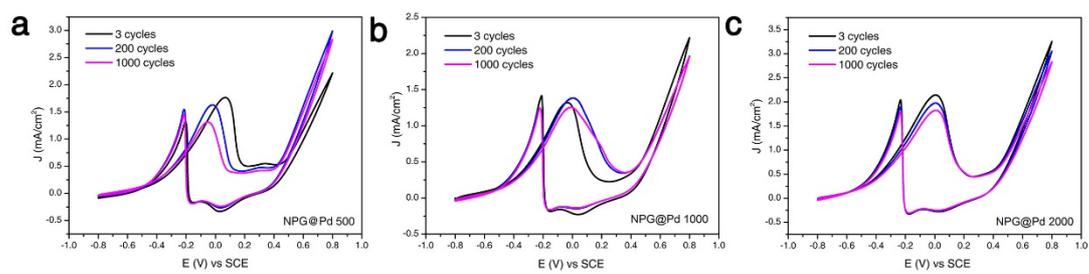

**Figure 7** Repetitive potential cycling stability test for sample (**a**) 'NPG@Pd 500' (**b**) 'NPG@Pd 1000' and (**c**) 'NPG@Pd 2000'.

*Yi Xu, Dr. Pak Man Yiu, Dr. Guangcun Shan,* Prof. Dr. Tamaki Shibayama, Prof. Dr. Seiichi Watanabe, Prof. Dr. Masato Ohnuma, Prof. Dr. Wei Huang,* Prof. Dr. Chan-Hung Shek**

**Fabrication of 3D Nanoporous Gold with Very Low Parting Limit Derived from Au-based Metallic Glass and Enhanced Methanol Electro-oxidation Catalytic Performance Induced by Metal Migration**

TOC

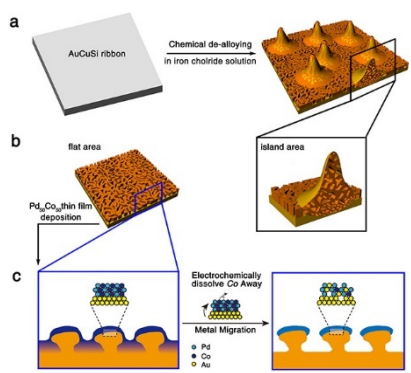